# Artificial Material with Negative Thermal Expansion: A Simple Geometrical Approach

Mikrajuddin Abdullah

[1)]*Department of Physics, Bandung Institute of Technology,*

[2)]*Research Center for Nanotechnlogy, Bandung Institute of Technology,*

*Jl. Ganesa 10 Bandung 40132, Indonesia*

*Email: mikrajuddin@gmail.com*

## Abstract

In the paper we report the modeling and design of material which has a negative thermal expansion (NTE). The basic assumption is a potential between the atoms in the material can be approximated by a Lennard-Jones potential (6-12) and the dominant interaction is only between nearest neighbors. We show that the material formed by alternating atomic layers wherein each layer contains a type of atom, and geometry of the arrangement of atoms is triangular, may experience an NTE when $\sigma_{AA} = \sigma_{AB} = \sigma_{BB}$, $\varepsilon_{AA} = \varepsilon_{BB} < \varepsilon_{AB}/4$ are satisfied, where $\sigma_{ij}$ and $\varepsilon_{ij}$ are Lennard-Jones parameters.

# 1. Introduction

In general, the material expands if the temperature increases. If the change in temperature is not too large, the change in the size satisfies $\Delta \psi = \beta \psi_0 \Delta T$, with $\psi$ is a variable stating the size of the material (length, area, or volume), $\psi_0$ is the initial size, $\Delta \psi$ is the size change due to temperature change $\Delta T$, and $\beta$ is the coefficient of thermal expansion associated with $\psi$ (thermal expansion coefficient of length, area, or volume).

The thermal expansion measured microscopically is the implications of the increase in the distance between atoms in the material. The atoms in the material are bound by a specified force and have certain equilibrium distances. The atoms always vibrate around the equilibrium positions. If the vibration displacement is very small then the potential experienced by the atoms around the equilibrium position can be approximated by harmonic potentials. The vibration around the equilibrium positions under the influence of harmonic potentials does not produce average displacements. However, if the displacement of atoms from the equilibrium position is large enough, the potential experienced by the atom around the equilibrium position is no longer a purely harmonious but slightly deviates from the harmonic form. The potential experienced by the atoms contains terms of power 3 or higher of the vibration displacement. As a result, the average displacement of atom from the equilibrium position is not zero. This resulted in an average distance of an atom changes and such changes depending on the temperature. This behavior led to expansion or contraction.

There have been many reports about the discovery of NTE materials. Initially, the materials with NTE properties were found in complex structures such as $ZrW_2O_8$ [1], $Zn_xCd_{1-x}(CN)_2$ [2], $Ag_3[Co(CN)_6]$ [3], $NbZr(PO4)_3$, $ZrV_2O_7$, $Sc_2(NO_4)_3$ [4], $Mn_3(Cu_{1-x}Ge_x)N$ [5], etc. For example, the coefficient of thermal expansion of $ZrW_2O_8$ is around $-2.6 \times 10^{-5}$ $K^{-1}$ [1]. Recently, materials with simple structures have also been observed to have NTE behavior such as CuO, $MnF_2$ [6], $ReO_3$ [7], and graphene [8]. The thermal expansion coefficient of CuO nanocrystal of a size about 5 nm is around $1.06 \times 10^{-4}$ $K^{-1}$ [6] and graphene has been observed to have a thermal expansion coefficient of around $-8.0 \pm 0.7 \times 10^{-6}$ $K^{-1}$ at temperatures between 200 K – 400 K [8].

The mechanisms which are considered to be responsible for the occurrence of NTE are the lattice vibrations, thermally-excited magnetic phase transitions, diffusion of ions or atoms in interstitial sites [9]. In nanoporous metals, twisting, rotation, and liberation are considered to be the cause of the NTE [10].

Studies on the properties of the NTE materials were more often conducted experimentally. The theory that underlies the occurrence of such mechanisms has not been widely discussed. Study with density functional theory has been reported by Lichtenstein [11] and the search for the pair potential model which gave rise to the phenomenon of the NTE was reported by Rechtman et al [12]. The purpose of this paper is to design an artificial material which has the NTE behavior using a fairly simple model. More specifically, this work was focused on materials that shrink at least in one direction if the temperature increases.

**Modeling and Discussion**

We can explain the mechanism of linear expansion with a simple approach as described by Kittel [13]. We will discuss specifically for solids. We will also limit the discussion to the case in which the interaction between the nearest neighboring atoms are very dominant and the next neighbors interaction can be ignored. We suppose that the potential between two atoms in the solid can be approximated with a Lennard-Jones potential (6-12) as follows [13]

$$U(x) = -4\varepsilon\left[\left(\frac{\sigma}{x}\right)^6 - \left(\frac{\sigma}{x}\right)^{12}\right] \quad (1)$$

where $x$ is the distance between atoms, $-\varepsilon$ is the depth of the potential valley measured from the potential when the atomic distance is infinite, and $\sigma$ is the distance between atoms when the potential is equal to zero. The distance between atoms in the potential valleys satisfies $dU/dx = 0$ and we can easily prove that distance is $x_0 = 2^{1/6}\sigma$.

Atoms always vibrate so that the distances between atoms always change. If the vibration displacement is not too large, we can approximate potential energy around the equilibrium position with the following polynomial

$$U(x) = U(x_0) + \frac{dU}{dx}\bigg|_{x_0} y + \frac{1}{2}\frac{d^2U}{dx^2}\bigg|_{x_0} y^2 + \frac{1}{6}\frac{d^3U}{dx^3}\bigg|_{x_0} y^3 + \frac{1}{24}\frac{d^4U}{dx^4}\bigg|_{x_0} y^4 \qquad (2)$$

where $y = x - x_0$. Since $x_0$ is the equilibrium distance, then $dU/dx|_{x_0} = 0$. Using the potential in equation (1) it is easy to prove that $U(x_0) = -\varepsilon$, $d^2U/dx^2|_{x_0} = 72\varepsilon/2^{1/3}\sigma^2$, $d^3U/dx^3|_{x_0} = -672\varepsilon/\sqrt{2}\sigma^3$, and $d^4U/dx^4|_{x_0} = 1512\varepsilon/2^{2/3}\sigma^4$. Furthermore, equation (2) can be rewritten as [13]

$$U(x) = U(x_0) + cy^2 - gy^3 - fy^4 \qquad (3)$$

with

$$c = \frac{1}{2}\frac{d^2U}{dx^2}\bigg|_{x_0} = \frac{36\varepsilon}{2^{1/3}\sigma^2} \qquad (4a)$$

$$g = -\frac{1}{6}\frac{d^3U}{dx^3}\bigg|_{x_0} = \frac{112\varepsilon}{\sqrt{2}\sigma^3} \qquad (4b)$$

$$f = -\frac{1}{24}\frac{d^4U}{dx^4}\bigg|_{x_0} = -\frac{63\varepsilon}{2^{2/3}\sigma^4} \qquad (4c)$$

Now we calculate the average displacement of atoms from the equilibrium position. We use the Maxwell-Boltzmann statistics to represent displacement distribution of atoms in order to obtain the average displacement [13]

$$\langle y \rangle = \frac{\int_{-\infty}^{+\infty} y \exp\left[\beta\left(-\varepsilon + cy^2 - gy^3 - fy^4\right)\right] dy}{\int_{-\infty}^{+\infty} \exp\left[\beta\left(-\varepsilon + cy^2 - gy^3 - fy^4\right)\right] dy} \tag{5}$$

with $\beta = -1/kT$ and $k$ is the Boltzmann constant. The factor $\exp[-\beta\varepsilon]$ in the numerator and denominator in equation (5) eliminate each other so that equation (5) becomes

$$\langle y \rangle = \frac{\int_{-\infty}^{+\infty} y \exp\left[\beta\left(cy^2 - gy^3 - fy^4\right)\right] dy}{\int_{-\infty}^{+\infty} \exp\left[\beta\left(cy^2 - gy^3 - fy^4\right)\right] dy} \tag{6}$$

The non-harmonic terms are generally very small, or $|gy^3 + fy^4| << 1$ so that we can make the following approximation: $\exp\left[-\beta(gy^3 + fy^4)\right] \approx 1 - \beta(gy^3 + fy^4)$. Thus, the numerator in the equation (6) can be approximated by

$$\int_{-\infty}^{+\infty} y \exp\left[\beta\left(cy^2 - gy^3 - fy^4\right)\right] dy \approx \int_{-\infty}^{+\infty} y \exp\left[\beta cy^2\right]\left[1 - \beta(gy^3 + fy^4)\right] dy$$

$$= \int_{-\infty}^{+\infty} y \exp\left[\beta cy^2\right] dy - \beta g \int_{-\infty}^{+\infty} y^4 \exp\left[\beta cy^2\right] dy - \beta f \int_{-\infty}^{+\infty} y^5 \exp\left[\beta cy^2\right] dy \tag{7}$$

Integrands in the first and the third terms in equation (7) are the odd functions so that the integral results are zero. Integral of the second term results $\beta g \int_{-\infty}^{+\infty} y^4 \exp\left[\beta cy^2\right] dy$ $= (3/4)\beta g \sqrt{\pi}/(-\beta c)^{5/2}$. In the integral in the denominator we can omit $gy^3 + fy^4$ so we may approximate $\int_{-\infty}^{+\infty} \exp[\beta cy^2] dy = \sqrt{-\pi/\beta c}$. Finally we get the average displacement of atoms or the change in the average distance between atoms becomes

$$\langle y \rangle = \frac{3g}{4c^2} kT = \frac{7}{2^{1/6} 108} \frac{\sigma}{\varepsilon} kT \tag{8}$$

It appears that the change the distance between atoms is strongly dependent on the depth of the potential energy valley between atoms. The deeper the valley, the change of the distance between atoms becomes smaller.

Consider a material in the form of a stack of layers of atoms A and B alternately as shown in Figure 1. At temperatures close to zero kelvin we assume the distance between atoms are all the same, say $a$. There are three types of bonds between nearest neighbor atoms ie A-A, A-B, and B-B bonds. The potential energies of each pair are as follows

$$U_{ij}(x) = -4\varepsilon_{ij} \left[ \left(\frac{\sigma_{ij}}{x}\right)^6 - \left(\frac{\sigma_{ij}}{x}\right)^{12} \right] \tag{9}$$

where i and j can be either atom A or B. Because at temperatures approaching zero distances between atoms are the same, ie $a$, we have $a = 2^{1/6} \sigma_{AA} = 2^{1/6} \sigma_{AB} = 2^{1/6} \sigma_{BB}$. Therefore

$$\sigma_{AA} = \sigma_{AB} = \sigma_{BB} = \sigma. \tag{10}$$

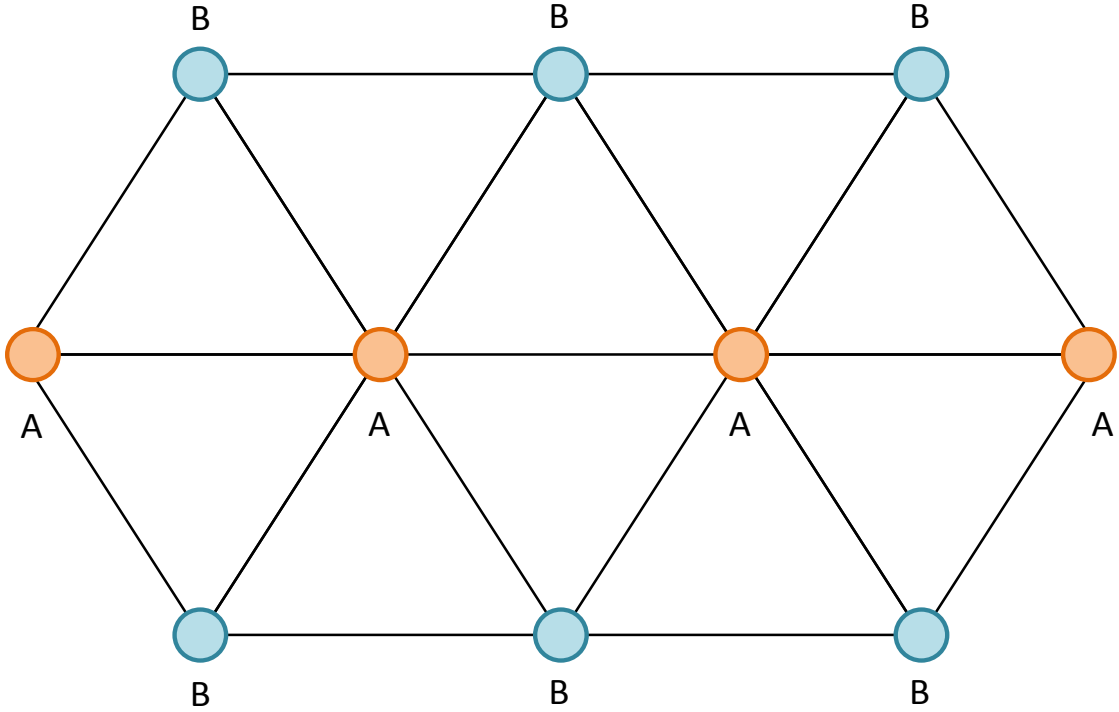

**Figure 1.** Material in the form of stack of layers of atoms A and B alternately. At temperatures approaching zero kelvin, the distances between the atoms are the same.

When the temperature is raised then the distances between atoms changes by

$$\langle y \rangle_{ij} = \frac{3g_{ij}}{4c_{ij}^2} kT$$

$$= \frac{7}{2^{1/6} 108} \frac{\sigma}{\varepsilon_{ij}} kT \qquad (11)$$

where i and j can be either A or B atoms. Look at Figure 2 that representes the geometry of the arrangement of atoms at temperatures close to zero kelvin and at arbitrary temperature *T*.

Suppose that at temperatures close to zero kelvin the transverse distance between the layers occupied by atoms A and B is $b$. Based on the figure it is clear that

$$b^2 = a^2 - \left(\frac{a}{2}\right)^2 = \frac{3}{4}a^2 \qquad (12)$$

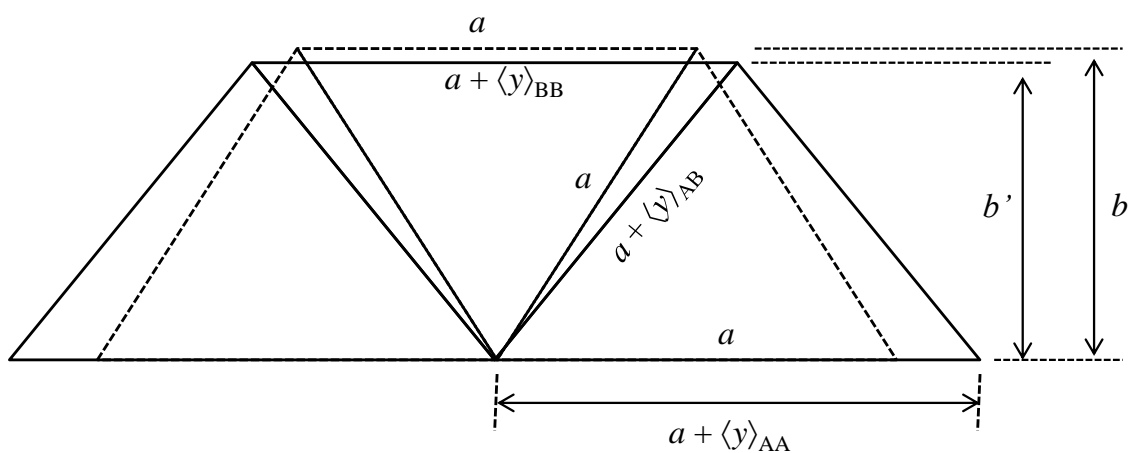

**Figure 2**. The geometry of the arrangement of atoms at temperatures close to zero kelvin (dashed lines) and at arbitrary temperature $T$ (solid lines).

When the temperature is raised, transverse distance between the layers occupied by atoms A and B changes to $b'$ satisfying

$$b'^2 = \left(a + \langle y \rangle_{AB}\right)^2 - \left(\frac{a + \langle y \rangle_{AA}}{2}\right)^2$$

$$= b^2 + 2a\langle y \rangle_{AB} + \langle y \rangle_{AB}^2 - \frac{1}{4}\left[2a\langle y \rangle_{AA} + \langle y \rangle_{AA}^2\right] \qquad (13)$$

Shrinkage in the transverse direction occurs when $b' < b$ or $b'^2 < b^2$ which implies

$$2a\langle y\rangle_{AB} + \langle y\rangle_{AB}^2 - \frac{1}{4}\left[2a\langle y\rangle_{AA} + \langle y\rangle_{AA}^2\right] < 0$$

The solution of this inequality is

$$\langle y\rangle_{AB} < -a + \sqrt{a^2 + \frac{1}{4}\left[2a\langle y\rangle_{AA} + \langle y\rangle_{AA}^2\right]}$$

Because generally $\langle y\rangle_{AA} \ll a$ then using the binomial approximation we obtain

$$\langle y\rangle_{AB} < \frac{1}{4}\langle y\rangle_{AA} + \frac{1}{8}\frac{\langle y\rangle_{AA}^2}{a}$$

If we ignore the quadratic term, the requirements for the transverse direction shrinkage to arise is $\langle y\rangle_{AB} < (1/4)\langle y\rangle_{AA}$. By using equation (11) we get

$$\varepsilon_{AB} > 4\varepsilon_{AA} \tag{14}$$

Next we examine the requirements for $\varepsilon_{BB}$. If we examine Figure 2, it is clear that changes in the horizontal distances of atoms in the layer occupied by atoms B must be exactly the same as changes in the horizontal distances of atoms in the layer occupied by atoms A, or $\langle y\rangle_{AA} = \langle y\rangle_{BB}$. This equality implies

$$\varepsilon_{BB} = \varepsilon_{AA} \tag{15}$$

Thus, to produce the transverse shrinkage, then the potential energy form should be as illustrated in Figure 3.

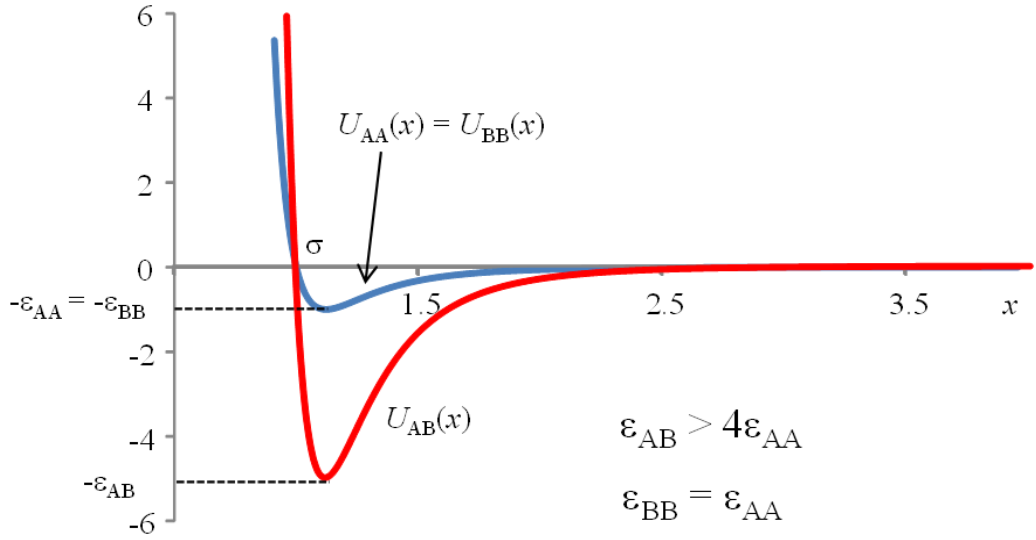

**Figure 3**. Illustration of the potential energy in order to produce the transverse shrinkage.

Now we calculate the transverse shrinkage as follows

$$\Delta z = b' - b$$

$$= \sqrt{b^2 + 2a\langle y\rangle_{AB} + \langle y\rangle^2_{AB} - \frac{1}{4}\left[2a\langle y\rangle_{AA} + \langle y\rangle^2_{AA}\right]} - b \qquad (16)$$

By using the binomial approximation we get

$$\Delta z \approx \frac{a}{b}\left[\langle y\rangle_{AB} - \frac{1}{4}\langle y\rangle_{AA}\right] = \frac{7}{162}2^{1/6}\sqrt{3}\left(\frac{1}{\varepsilon_{AB}} - \frac{1}{4\varepsilon_{AA}}\right)\sigma kT \qquad (17)$$

Because the requirements for shrinkage to arise is $\varepsilon_{AB} > 4\varepsilon_{AA}$, as expressed by equation (14), then we get $\Delta z < 0$.

From the description above, we conclude that if the material is composed by two types of atoms with the geometry as shown in Figure 1 and parameters for the Lennard-Jones potential energy satisfies the equation (10), (14) and (15) then the material experiences expansion in the direction connecting the same atoms and experiences shrinkage in the perpendicular direction (transversal). In the special case when $\varepsilon_{AB} = 4\varepsilon_{AA}$ then we get $\Delta z = 0$. This means that the material does not experience thermal expansion in the transverse direction although the temperature changes. Synthesis of material with zero expansion coefficient has been reported by Salvador et al [14] on material YbGaGe. In principle, this material combines two types of material, one has a positive thermal expansion coefficient and the other has a negative thermal expansion coefficient.

**Conclusion**

We have shown using a fairly simple method that material with the structure shown in Figure 1 can have NTE in the transverse direction as long as the interactions between atoms are only dominant for nearest neighbor and assumed satisfy the Lennard-Jones potential 6-12 with the parameters given by equation (10), (14) and (15). Basically the distances between the atoms all increase when the temperature increases, but due to the geometry shown in Figure 1 then the transverse distance between atomic layers can decrease if the temperature increases.

**Acknowledgement**

This work was supported by a research grant (No. 310y/I1.C01/PL/2015) from the Ministry of Research and Higher Education, Republic of Indonesia, 2015-2017.